\documentclass[
aps,prd,
12pt,
nopreprintnumbers,
showpacs,
eqsecnum,
nofootinbib
]{revtex4-1}

\def\Tr{{\rm Tr\,}}

\begin{document}

\title{
Equations of Motion in Double
Field Theory:\\
From Particles to Scale Factors}
\author{Nahomi Kan}\email[]{kan@yamaguchi-jc.ac.jp}
\affiliation{
Yamaguchi Junior College,
Hofu-shi, Yamaguchi 747--1232, Japan}
\author{Koichiro Kobayashi}\email[]{m004wa@yamaguchi-u.ac.jp}
\author{Kiyoshi Shiraishi}\email[]{shiraish@yamaguchi-u.ac.jp}
\affiliation{
Yamaguchi University,
Yamaguchi-shi, Yamaguchi 753--8512, Japan}
\date{\today}

\begin{abstract}
In double field
theory, the equation of motion for a point particle in the background field is
considered. We find that the motion is described by a geodesic flow in the doubled
geometry. Inspired by the analysis on the particle motion, we consider a modified
model of quantum string cosmology, which includes two scale factors.
\end{abstract}


\maketitle

\section{Introduction}

T-duality is an important symmetry in string theory~\cite{R}. 
This symmetry can be interpreted as an invariance under the interchanging
of coordinates and dual coordinates. 
Dual coordinates are also reported to be important in string 
field theory (SFT)~\cite{0}. 

Recently, Hull, Zwiebach, and Hohm constructed the theory
of the massless field with a higher symmetry of spacetime including
dual coordinates; This is referred to as double field theory (DFT). The original
work is given in \cite{1}, and later developments can be found in
\cite{2,3,4,5,H,H2}.%
\footnote{Other developments that we were not aware of during the completion
of the main crux of this study are given in
\cite{HK2,Cop,Thompson,HKZ,ADKL}.}
Through this theory, Hull, Zwiebach, and Hohm
clarified the T-duality symmetry of the massless field, new algebra, and the symmetry
related to the theory.

Because T-duality was discovered and interpreted as an exchanging symmetry of
the Kaluza-Klein (KK) modes and winding modes in compact space, we are simply
interested in the extent to which the theory of zero mode field alone
can describe the special features of string theory. 
More recently Jeon, Lee, and Park proposed the construction of the Yang-Mills theory
as DFT
\cite{JLP2} using their projection-compatible differential geometrical methods in
\cite{4}. It is interesting to study the duality-symmetric extension of various field
theories.

In the present paper, the formalism in \cite{4}, in which the
geometrical aspect of DFT is introduced is concisely reviewed, and subsequently the 
couplings of zero-mode fields and a particle are examined along with the equation of
motion for the particle in the background fields. 

One of our motivations for this investigation is the
importance of clarifying the meaning of geodesics in space with consideration of
dual coordinates.
In addition, since the origin of the
generalized metric is considered for a string, we believe that the
coupling to the background fields can be neatly interpreted for particles
as well.

In the next section, we review the projection-compatible approach to DFT \cite{4}. 
In
Sec.~\ref{ems}, the equation of motion for a string, as proposed by Duff \cite{D2} is
reviewed.  After these reviews, we attempt to find the geometrical meaning of particle
motion in Sec.~\ref{emp}.  In Sec.~\ref{pgf}, we analyze the particle motion in the
case of Hamiltonian formalism. The equation of  geodesic flow is applied to the KK
type model in Sec.~\ref{KKbf}. The quantization of particles in the DFT background is
studied in Sec.~\ref{part}. The KK mass spectrum of the scalar field is also
investigated. Inspired by the above approach, we propose a modified model of
DFT for cosmological evolution equations in Sec.~\ref{ccos}. In Sec.~\ref{ms},
the mini-superspace formulation of our model is explained, and the
quantum cosmological aspects of our model are examined in Sec.~\ref{QC}. The final
section is devoted to the summary and prospects.

\section{review of projection-compatible approach
\label{pca}}
The constant metric is assumed to be expressed as the following $2D \times 2D$ matrix,
\begin{equation}
\eta_{AB}=\left(
\begin{array}{cc}
0 & \delta^{a}{}_{\nu}\\
\delta_{\mu}{}^{b}\ & 0
\end{array}\right)\,.
\end{equation}
Here, the suffixes $A, B,\ldots$ range over $1, 2,\ldots, 2D$,
while $\mu, \nu,\ldots$ as well as $a, b,\ldots$ range over $1, 2,\ldots, D$.
The suffixes are entirely raised and lowered by this constant metric.
Of course,
\begin{equation}
\eta^{AC}\eta_{CB}=\delta^{A}{}_B\,,
\end{equation}
is satisfied.

The generalized metric is defined as follows. 
\begin{equation}
{\cal H}_{AB}=\left(
\begin{array}{cc}
g^{ab} & ~~-g^{a\sigma}b_{\sigma\nu}\\
b_{\mu\sigma}g^{\sigma b} & ~~~g_{\mu\nu}-b_{\mu\rho}
g^{\rho\sigma}b_{\sigma\nu}
\end{array}\right)\,.
\end{equation}
Here, $g_{\mu\nu}$ and $b_{\mu\nu}$ are the metric in $D$ dimensions and the
antisymmetric tensor, respectively. It should be noted that the inverse of the
generalized metric, or the matrix  ${\cal H}^{AB}$ satisfying
\begin{equation}
{\cal H}^{AC}{\cal H}_{CB}=\delta^{A}{}_B\,,
\end{equation}
is obtained by raising the suffixes by
the constant metric, as
\begin{equation}
{\cal H}^{AB}=
\eta^{AC}{\cal H}_{CD}\eta^{DB}=
\left(
\begin{array}{cc}
g_{ab}-b_{a\rho}g^{\rho\sigma}b_{\sigma b} &
~~~b_{a\sigma}g^{\sigma \nu}\\
-g^{\mu\sigma}b_{\sigma b} & ~~~g^{\mu\nu} 
\end{array}\right)\,.
\end{equation}

The following projection matrices are defined on the basis of the existence of two
kinds of metrics.
\begin{equation}
P^A{}_B\equiv\frac{1}{2}(\delta^A{}_{B}+{\cal H}^A{}_B)\,,\quad
\bar{P}^A{}_B\equiv\frac{1}{2}(\delta^A{}_{B}-{\cal H}^A{}_B)\,,
\end{equation}
which satisfy
\begin{equation}
P^2=P\,,\quad \bar{P}^2=\bar{P}\,,\quad P\bar{P}=\bar{P}P=0\,.
\end{equation}
From this, one can derive the identities
\begin{equation}
P(\partial_A P)P=\bar{P}(\partial_A \bar{P})\bar{P}=0\quad{\rm or}\quad
P_{D}{}^{B}(\partial_A{\cal H}_{BC})P^{C}{}_{E}=\bar{P}_{D}{}^{B}(\partial_A{\cal
H}_{BC})\bar{P}^{C}{}_{E}=0\,.
\end{equation}

Now, the projection-compatible derivative is defined. In other words, both the metrics
are ``covariantly constant,'' {i.e.},
\begin{equation}
\nabla_A \eta_{BC}=0\,,\qquad\nabla_A{\cal H}_{BC}=0\,.
\end{equation}
So, the covariant derivative of the projection of an arbitrary tensor coincides with
the projection of the covariant derivative of the tensor.  Jeon {et al.}\cite{4} found
that the covariant derivatives including the following connection have the
character,
\begin{equation}
\Gamma_{ABC}\equiv
2P_{[A}{}^{D}\bar{P}_{B]}{}^{E}\partial_C
P_{DE}+2(\bar{P}_{[A}{}^{D}\bar{P}_{B]}{}^{E}-P_{[A}{}^{D}P_{B]}{}^{E})\partial_D
P_{EC}\,.
\end{equation}

They also obtained the action for the generalized metric, which was previously found
by Hohm, Hull, and Zwiebach \cite{1,2,3,5,H,H2}
\begin{eqnarray}
S&=&\int dx d\tilde{x}\, e^{-2d}\left(\frac{1}{8}{\cal
H}^{AB}\partial_A{\cal H}^{CD}\partial_B{\cal H}_{CD}-\frac{1}{2}{\cal
H}^{AB}\partial_B{\cal H}^{CD}\partial_D{\cal H}_{AC}\right.\nonumber \\
& &\qquad\qquad\qquad\left.-2\partial_A d\partial_B{\cal H}^{AB}+4{\cal
H}^{AB}\partial_A d\partial_B d{}\right)\,,
\label{ga}
\end{eqnarray}
from the consideration of the projection-compatible geometrical quantities.
Here, $e^{-2d} = \sqrt{-g}\,e^{-2\phi}$ and $\phi$ is the dilaton field.  If we set
all the derivatives on the fields with respect to the dual coordinate zero
($\tilde{\partial}^a=0$), the action for the well-known effective theory for
the zero-mode (massless) field in string theory is obtained as 
\begin{equation}
S=\int dx
\sqrt{-g}\,e^{-2\phi}\left[R+4(\partial\phi)^2-\frac{1}{12}H^2\right]\,,
\end{equation}
where 
the three-form field $H=db$ is the field
strength of the Kalb-Ramond 2-form $b_{ij}$.

In a cosmological context, we consider that all the fields depend only on time
coordinate $t$ and assume that the metric and the antisymmetric tensor in $D+1$
dimensions are respectively,
\begin{equation}
g_{\mu\nu}=\left(
\begin{array}{cc}
-1 & 0\\
0 & G_{\mu\nu}(t)
\end{array}\right)\,,\qquad
b_{\mu\nu}=\left(
\begin{array}{cc}
0 & 0\\
0 & B_{\mu\nu}(t)
\end{array}\right)\,.
\end{equation}
Then, the action given in (\ref{ga}) is reduced to
\begin{equation}
S\rightarrow -\int dt\, e^{-\Phi}\left(\frac{1}{8}
\Tr(\dot{M}\eta\dot{M}\eta)+\dot{\Phi}^2\right)\,,
\end{equation}
where the dot represents the time derivative, $\Phi\equiv 2d$, and
\begin{equation}
M\equiv\left(
\begin{array}{cc}
G^{-1} & ~~-G^{-1}B\\
BG^{-1} & ~~~G-BG^{-1}B
\end{array}\right)\,.
\end{equation}
It is also assumed that $\eta$ is a $2D \times 2D$ matrix. 
The action exactly corresponds to the one considered in the string
cosmology and which has been studied by several authors \cite{GV}.

\section{the equation of motion for a string
\label{ems}}
In this section, let us review the work of Duff~\cite{D2}. 
He investigated the equation of motion and
the duality in string theory.  In the next section, we will study the equation
of motion of a particle in DFT on the basis of the consideration in the
present section. Although the extent of the remaining influences of the string duality
to the motion of particles is known, the consideration on strings and particles in the
background fields is indispensable because ``the geometry of spacetime tells particles
how to move.''%
\footnote{The original sentence is found in \cite{MTW}.}
After all, we are interested in various geometrical characters in DFT.

Now, let us turn to the review of the paper of Duff~\cite{D2}. Coordinates are
combined
with dual coordinates to be
\begin{equation}
X^A\equiv\left(
\begin{array}{c}
\tilde{x}_a\\ x^{\mu}
\end{array}\right)\,,
\end{equation}
which is an $O(D,D)$ vector. 

For simplicity we consider the standard line element on the world sheet,
$\eta_{ij} d\xi^id\xi^j=d\tau^2-d\sigma^2$.  At this time,
Duff's equation of motion is \cite{D2}
\begin{equation}
\partial^i({\cal H}_{AB}\partial_iX^B)=0\,,
\label{sa}
\end{equation}
where the notation is renewed.

The equation of duality, or the ``BPS'' equation providing the solution of this
equation of motion is given by \cite{D2}
\begin{equation}
\eta_{AB}{X'}^B={\cal H}_{AB}\dot{X}^B\,,
\end{equation}
where the prime ($'$) and the dot ($\dot{~}$) denote the derivatives with respect to
$\sigma$ and
$\tau$, respectively.

This equation is equivalent to
\begin{equation}
P_{AB}\frac{\partial X^B}{\partial\sigma^{-}}=0\,,\qquad
\bar{P}_{AB}\frac{\partial X^B}{\partial\sigma^{+}}=0\,,
\end{equation}
where $\sigma^{\pm}=\tau\pm\sigma$. 
Then, the equation can be expressed in terms of $X_L^B(\sigma^{+})$ and
$X_R^B(\sigma^{-})$, each of which is the function of their respective coordinates, as
\begin{equation}
P^{A}{}_{B}X^B=X_L^A\,,\qquad \bar{P}^{A}{}_{B}X^B=X_R^A\,,
\end{equation}
in the fixed background fields. As is well known,
the string coordinate and its dual are written in $X=X_L+X_R$ and
$\tilde{X}=X_L-X_R$, respectively, and hence, 
the relation of $X\sim\tilde{X}$ will show that the spacetime described by
the set of coordinates is the projected space obtained with
the projection matrix $P$.
This is consistent with the projection-compatible procedure of Jeon {et al.}

The relation between the duality and the projection in the equation of the string has
been understood above.  Duff explicitly claimed in his paper \cite{D2} that the
duality (rotation) is not the symmetry of the action but the symmetry of the
equation of motion of the string.  In other
words, the equation of motion does not necessarily originate from an action with
duality invariance. This certainly appears to be the case; the equation of motion
(\ref{sa}) does not have the action in most general background fields.

\section{Is the equation of motion for a particle the geodesic equation?
\label{emp}}
Next, we consider the equation of motion for a particle.
Let us start with the context of the differential geometry with projection.  For a
usual common aspect of general relativity, the geodesic equation is given by the
following expression \cite{Wald}
\begin{equation}
U^\mu\nabla_\mu U^\nu=0\,,
\end{equation}
where $U^\mu\equiv \frac{dx^\mu}{ds}=\dot{x}^\mu$, $s$ being a parameter.

The corresponding equation in the projection compatible geometry of Jeon {et al.}
is considered to be 
\begin{equation}
U^A\nabla_A U^B=U^A(\partial_A U^B+\Gamma_A{}^B{}_CU^C)=0\,,
\label{deq}
\end{equation}
where $U^A=(\tilde{U}_a, U^{\mu})^T=\frac{d{X}^A}{ds}$.

We explicitly calculate the connection
$\Gamma_A{}^B{}_C$ for $b_{\mu\nu} =0$.  The suffixes $\mu$
and $\nu$ indicate those of the usual coordinates while  $a$ and $b$ are
those of dual
coordinates. Note that both suffixes can stick to the single metric $g$. Either letter
ranges over $1$ to $D$,  
which is considered carefully in the sum by the Einstein rule.  Moreover,
the derivative with respect to the dual coordinates is denoted by
$\tilde{\partial}$. The elements of $\Gamma_A{}^B{}_C$ are found to be
\begin{eqnarray}
\Gamma_{\nu}{}^{\mu}{}_{\lambda}&=&
\left\{{}_\nu{}^\mu{}_\lambda\right\}
-\frac{1}{2}g_{\lambda\sigma}
\partial_{\nu}g^{\sigma\mu}\,,\\
\Gamma_{\nu}{}^{\mu
c}&=&\frac{1}{2}\left(g^{\mu\sigma}\tilde{\partial}^cg_{\sigma\nu}
-g^{cd}\tilde{\partial}^{\mu}g_{d\nu}\right)\,,\\
\Gamma^{a\mu}{}_{\lambda}&=&-
\left\{{}^a{}_\lambda{}^\mu\right\}
+\frac{1}{2}g^{\mu\sigma}\tilde{\partial}^a
g_{\sigma\lambda}\,,\\
\Gamma^{a\mu
b}&=&\frac{1}{2}\left(g^{b\sigma}{\partial}_{\sigma}g^{\mu a}
-g^{\mu\sigma}{\partial}_{\sigma}g^{ab}\right)\,,\\
\Gamma^{b}{}_{a}{}^{c}&=&\left\{{}^b{}_a{}^c\right\}
-\frac{1}{2}g^{cd}\tilde{\partial}^b
g_{da}\,,\\
\Gamma^{b}{}_{a\lambda}&=&\frac{1}{2}\left(g_{ad}{\partial}_{\lambda}
g^{db} -g_{\lambda d}{\partial}_{a}g^{db}\right)\,,\\
\Gamma_{\nu a}{}^{b}&=&-
\left\{{}_\nu{}^b{}_a\right\}
+\frac{1}{2}g_{ad}
\partial_{\nu}g^{db}\,,\\
\Gamma_{\nu
a \lambda}&=&\frac{1}{2}\left(g_{\lambda b}\tilde{\partial}^bg_{a\nu}
-g_{ab}\tilde{\partial}^bg_{\nu\lambda}\right)\,,
\end{eqnarray}
where
\begin{eqnarray}
\left\{{}_\nu{}^\mu{}_\lambda\right\}
&\equiv&\frac{1}{2}g^{\mu\sigma}\left(
\partial_\nu g_{\sigma\lambda}+\partial_\lambda g_{\sigma\nu}-
\partial_\sigma g_{\nu\lambda}\right)\,,\\
\left\{{}^b{}_a{}^c\right\}
&\equiv&\frac{1}{2}g_{ad}\left(
\tilde{\partial}^b g^{dc}+\tilde{\partial}^c g^{db}-
\tilde{\partial}^d g^{bc}\right)\,.
\end{eqnarray}

From the project space ansatz $\bar{P}U=0$, we are forced to use
$\tilde{U}_a=g_{a\nu}U^\nu$. Moreover, we set $\tilde{\partial}g=0$ as in the
interpretation of DFT. Next, we find that the equation (\ref{deq}) reads
\begin{equation}
U^\mu\partial_\mu U^\nu+\frac{1}{2}g^{\nu\mu}(\partial_{\rho}g_{\mu\sigma}+
\partial_{\sigma}g_{\mu\rho})U^\rho U^\sigma=0\,.
\end{equation}
This equation is equivalent to $\frac{d}{ds}(g_{\mu\sigma}\dot{x}^\sigma)$,
which is similar to the equation of motion for
the string considered by Duff. However,
it is obviously different from the usual geodesic equation in general relativity
(or differential geometry).
It seems that the physical interpretation of the equation is
difficult.  If the equation is true, the Newtonian gravity cannot be obtained by
considering the ``Newtonian limit.''

In general, it is understood that the usage of the projection has a problem. 
From brief calculation, the following equation can be shown, 
\begin{equation}
P_{D}{}^{A}\Gamma_{(A}{}^{B}{}_{C)}P^{C}{}_E=
P_{D}{}^{A}\left[\frac{1}{2}{\cal H}^{BF}\left(\partial_{A}{\cal
H}_{FC}+\partial_{C}{\cal
H}_{FA}\right)\right]P^{C}{}_E\,.
\end{equation}
It is interesting that because $P\partial_A{\cal H} P=0$, we can show that
\begin{equation}
P_{D}{}^{A}\Gamma_{(A}{}^{B}{}_{C)}P^{C}{}_E=
P_{D}{}^{A}\left[\frac{1}{2}{\cal H}^{BF}\left(\partial_{A}{\cal
H}_{FC}+\partial_{C}{\cal
H}_{FA}-\partial_{F}{\cal H}_{AC}\right)\right]P^{C}{}_E
\end{equation}
is also true.
In this case, the equation is considered to be the one derived from the
Lagrangian
\begin{equation}
L=\frac{1}{2}{\cal H}_{AB}\dot{X}^A\dot{X}^B\,,
\end{equation}
with the projection such that the condition $\bar{P} U=0$ is imposed. 

No physical interpretation was obtained from (\ref{deq}). 
Since  there was obviously no physical requirement in the starting point,
we can correct the equation for a particle.
However, a clear geometrical point of view is missing.

Although the projection-compatible method is still valid for the field theory,
we noticed that the projection of the motion of the particle (or string) should be
handled as a constraint. In the next section, we consider the
motion of a particle in the spacetime described by the generalized metric in the
analytical mechanics of the constraint system.

\section{Projection and geodesic flow
\label{pgf}}
We treat the constraint using the Lagrange multiplier. 
The following Lagrangian is adopted, and the mechanics derived from it are
considered.
\begin{equation}
L=\frac{1}{2}{\cal H}_{AB}\dot{X}^A\dot{X}^B+\lambda^A\bar{P}_{AB}\dot{X}^B\,.
\end{equation}
Here, $\lambda^A$ is an undecided multiplier. 
The Euler-Lagrange equation leads to the constraint $\bar{P}\dot{X}=0$,
which is adopted as the projection in the previous section.

We find that the conjugate momentum of $X^A$ is
\begin{equation}
p_A=\frac{\partial L}{\partial \dot{X}^A}={\cal
H}_{AB}\dot{X}^B+\lambda^B\bar{P}_{BA}\,,
\end{equation}
and
the conjugate momentum of $\lambda^A$ is
\begin{equation}
\pi_A=\frac{\partial L}{\partial \dot{\lambda}^A}=0\,.
\end{equation}
Then, the Hamiltonian is defined as
\begin{equation}
H=p_A\dot{X}^A+\pi_A\dot{\lambda}^A-L=
\frac{1}{2}{\cal
H}^{AB}(p_A-\lambda^C\bar{P}_{CA}) (p_B-\lambda^D\bar{P}_{DB})\,.
\label{oldH}
\end{equation}

The equation for the conjugate momentum of the multiplier
$\lambda^A$ becomes
\begin{equation}
\dot{\pi}_A=
-\frac{\partial H}{\partial \lambda^A}
=\bar{P}_{AB}{\cal H}^{BC}(p_C-\lambda^D\bar{P}_{DC})=
-\bar{P}_{AB}(p^B-\lambda^B)\,,
\end{equation}
which will vanish. The multiplier can be determined from the above equation as 
\begin{equation}
\lambda_A=p_A+P_{AB}M^B\,,
\end{equation}
where $M^B$ is an arbitrary vector. 
When this solution is substituted into the Hamiltonian (\ref{oldH}), we obtain a new
Hamiltonian 
\begin{equation}
H_{\star}=
\frac{1}{2}{\cal
H}^{AB}P_{AC}p^C P_{BD}p^D=\frac{1}{2}P^{AB}p_A p_B\,,
\end{equation}
where the arbitrariness in the solution has disappeared. 
If we take $p_A=(\tilde{p}^a, p_\mu)$, the new Hamiltonian can be rewritten as
\begin{equation}
H_\star=\frac{1}{2}\tilde{p}^a p_a+
\frac{1}{4}g^{\mu\nu}(p_\mu-b_{\mu\rho}\tilde{p}^\rho)(p_\nu-b_{\nu\sigma}
\tilde{p}^\sigma)
+\frac{1}{4}g_{ab}\tilde{p}^a\tilde{p}^b\,.
\end{equation}

Using the new Hamiltonian, we obtain
\begin{equation}
\dot{X}^A=\frac{\partial H_\star}{\partial p_A}=P^{AB}p_B\,.
\label{jk}
\end{equation}
This clearly satisfies the constraint $\bar{P}\dot{X}=0$.  
Furthermore, we can obtain the equation of motion 
\begin{equation}
\dot{p}_A=-\frac{\partial H_\star}{\partial X^A}=-\frac{1}{2}\partial_A
P^{BC}p_B p_C=-\frac{1}{4}\partial_A
{\cal H}^{BC}p_B p_C\,.
\label{peq}
\end{equation}
These equations describe the geodesic flow in the system. Since equation
(\ref{jk}) is not always solvable for $p_A$, rewriting it a generalized
second-rank differential equation for $X^A$ is not possible.  That is, the expression
for the second derivative of $X^A$ becomes
\begin{eqnarray}
\ddot{X}^A&=&\frac{d}{ds}(P^{AB}p_B)=\dot{X}^C(\partial_CP^{AB})p_B+
P^{AB}\dot{p}_B\nonumber \\
&=&\dot{X}^C(\partial_CP^{AB})p_B-\frac{1}{2}
P^{AB}\partial_BP^{CD}p_Cp_D\,,
\label{gd}
\end{eqnarray}
where the momentum remains unsolved in general. 

Now, let us take the condition $\tilde{\partial}^a=0$ for the correspondence with
DFT.  In this case, the $D$-dimensional fixed vector ${\tilde{p}}^a={\tilde{p}} ^a_0$ 
(remembering that $p_A=(\tilde{p}^a, p_\mu)$) is a
general solution, because $\dot{\tilde{p}}^a=0$ can be obtained from equation
(\ref{peq}). 

In this section, we consider $\tilde{p}^a_0=0$ (for all dual coordinate components) to
assess the interpretation of the equation.  In this case, the equation (\ref{jk})
leads to
$\dot{x}^\mu=\frac{1}{2} g^{\mu\nu} p_{\nu} $.
Namely, $p_\mu=2g_{\mu\nu}
\dot{x}^\nu$ is obtained, and we substitute it into (\ref{gd}) and obtain
\begin{eqnarray}
\ddot{x}^\mu&=&\dot{x}^\nu(\partial_\nu g^{\mu \rho})g_{\rho\sigma}\dot{x}^\sigma
-\frac{1}{2}
g^{\mu\rho}(\partial_\rho g^{\sigma\nu})g_{\sigma\lambda}\,\dot{x}^\lambda
g_{\nu\tau}\dot{x}^\tau\nonumber \\
&=&-\frac{1}{2}g^{\mu\nu}\left(\partial_\rho g_{\nu\sigma}+\partial_\sigma g_{\nu\rho}-
\partial_\nu g_{\rho\sigma}\right)\dot{x}^\rho\dot{x}^\sigma\,.
\end{eqnarray}
This equation is nothing but $\ddot{x}^\mu+\left\{{}_{\rho\sigma}^{\,\mu\,}\right\}
\dot{x}^\rho\dot{x}^\sigma=0$, and the geodesic equation in a usual $D$ dimensional
spacetime.  It should be noted that the the geodesic equation has been obtained under
the condition $\tilde{p}_a=0$, with and without the existence of
$b_{\mu\nu}$.

We have obtained the geodesic equation in the $D$-dimensional spacetime from the
geodesic flow in the $2D$-dimensional space described by the generalized metric with
natural assumptions. In the next section, we will show the dynamical equation
with non-zero $\tilde{p}^a_0$.

\section{Kaluza-Klein background fields
\label{KKbf}}
In this section, we assume the factorized background fields, namely, the KK
background fields. Because the signature definition of the $D$-dimensional metric is
not essential here, it is assumed $\mu=1, 2, 3, \dots, D$. We consider the KK
independent of the {\it first} coordinate, {i.e.},
$\partial_1=0$.  From equation (\ref{peq}),
$p_1=constant$ becomes the solution in this case.  Moreover, as in the previous
section, 
$\tilde{\partial}^a=0$ is assumed and, therefore, $\tilde{p}^a$ is a constant vector.
Here, we suppose that
$\tilde{p}^i=0$ ($i=2,3,\dots, D$) while $\tilde{p}^1=constant$. 

Taking all the assumptions into account, we rewrite the Hamiltonian as
\begin{eqnarray}
H_\star(\{
x^i, p_i\}
)&=&\frac{1}{2}\tilde{p}^1 p_1+
\frac{1}{4}g^{ij}(p_i-\tilde{p}^1 B_i)(p_j-\tilde{p}^1 B_j)\nonumber \\
& &+\frac{1}{2}g^{i1}p_1(p_i-\tilde{p}^1 B_i)
+\frac{1}{4}g^{11}p_1^2
+\frac{1}{4}g_{11}(\tilde{p}^1)^2\,,
\end{eqnarray} 
where $B_i\equiv b_{i1}$. 

In addition, we now consider the KK 
ansatz; for this, the $D$-dimensional metric
can be decomposed as
\begin{equation}
g_{\mu\nu}=\left(
\begin{array}{cc}
R^2 & R^2A_i   \\
R^2A_j~ &\hat{g}_{ij}+R^2A_iA_j  
\end{array}
\right)\,,\quad
g^{\mu\nu}=\left(
\begin{array}{cc}
R^{-2}+A_k A_\ell \hat{g}^{k\ell} & ~-A_\ell \hat{g}^{i\ell}  \\
 -A_k \hat{g}^{kj}  & \hat{g}^{ij} 
\end{array}
\right)\,,
\end{equation}
where $R$ (constant) is interpreted as the length scale of the {\it first} dimension.

From all the statements above, we finally obtain
\begin{equation}
H_\star(\{x^i, p_i\})=
\frac{1}{4}\hat{g}^{ij}(p_i-p_1 A_i-\tilde{p}^1 B_i)(p_j-p_1 A_j-\tilde{p}^1 B_j)
+\frac{1}{4}\left(\frac{1}{R}p_1+R\tilde{p}^1\right)^2\,.
\end{equation}
At this time, we immediately obtain
\begin{equation}
\dot{x}^i=\frac{\partial H_\star}{\partial p_i}=
\frac{1}{2}\hat{g}^{ij}(p_j-p_1 A_j-\tilde{p}^1 B_j)\,.
\end{equation}
Substituting this into $\dot{p}_i=-\frac{\partial H_\star}{\partial x^i}$ 
and eliminating $p_i$ and their derivatives, we have the equation of motion
\begin{equation}
\ddot{x}^i+\hat{\Gamma}^i_{jk}\dot{x}^j\dot{x}^k=\hat{g}^{ij}{\cal F}_{jk}\dot{x}^k
\,,
\end{equation}
where $\hat{\Gamma}^i_{jk}\equiv \frac{1}{2}\hat{g}^{i\ell}(\partial_j \hat{g}_{\ell
k}+
\partial_k \hat{g}_{\ell j}-\partial_\ell \hat{g}_{jk})$ and
${\cal F}_{jk}=\partial_j \frac{1}{2}(p_1 A+\tilde{p}^1 B)_k-\partial_k
 \frac{1}{2}(p_1 A+\tilde{p}^1 B)_j$.

It can be understood that two $U(1)$ gauge fields have a different charge coupled to
the particle in the KK 
background field. From the classical theory,
the quantization of the coupling constant cannot be decided.  

We have found the equation of motion for the doubly charged particle in the constraint
system  in the doubled spacetime. Incidentally,
the equation of motion for the particle limit of the string has also been considered,
by Jafarizadeh and Rezaei-Aghdam \cite{JR}.

\section{quantization of particles
\label{part}}
Here, we consider the problem of the quantization of the particle. 
One of the effective approaches to quantization is the method of 
phase space action \cite{Pavsic}. 
Let us consider the following action
\begin{equation}
I=\int ds \left[p_A \dot{X}^A-\frac{N}{2}\left(P^{AB}p_A p_B+\frac{1}{2}m^2
\right)\right]\,,
\end{equation}
where the dot indicates the derivative with respect to $s$. 
$N$ is a Lagrange multiplier.  The constant $m$ is
allowed under the symmetry and is stated below.
If we consider the variation with respect to $p_A$, we obtain
$\frac{1}{N}\frac{d{X}^A}{ds}= P^{AB} p_B$, whereas if we consider the variation with
respect to $X^A$, we obtain
$\frac{1}{N}\frac{d{p}^A}{ds}+\frac{1}{2} \partial_A P^{BC} p_Bp_C=0$.  
This action is invariant if the parameter $s$ is changed, provided that 
$N$ cancels the variation (that is, $N$ acts as an einbein).  Therefore, the
Euler-Lagrange equations become equivalent to the expressions for the equation of
motion in the previous section if $N=1$ is assumed as a gauge choice.

When we formulate a Hamiltonian again from this phase space Lagrangian, we find
\begin{equation}
H_{\flat}=\frac{N}{2}\left(P^{AB}p_A p_B+\frac{1}{2}m^2\right)\,.
\end{equation}
From the condition $\frac{\partial H}{\partial N} \approx 0$, we obtain
$H_\flat|_{N=1}
\approx 0$. If the quantum state satisfies this assumption, the constraint
condition is expressed as
\begin{equation}
\left(P^{AB}p_A p_B+\frac{1}{2}m^2\right)\varphi=0\,,
\end{equation}
where $\varphi$ can be recognized as a scalar wave function.
Then, the wave equation of Klein-Gordon type is obtained, if we replace 
$p_A\rightarrow -i\partial_A$ here with an arbitrary ordering of the
operator. 
We can estimate
the second-quantized action of a scalar field as
\begin{equation}
S_\flat\stackrel{}{=}\int dx d\tilde{x}\, e^{-2d} \left(-P^{AB}\partial_A\varphi 
\partial_B\varphi-
\frac{1}{2}m^2\varphi^2\right)\,,
\end{equation}
where $\varphi$ is a real scalar field.

Now, we incorporate the KK ansatz and the assumption adopted in the previous section.
In addition, we assume
\begin{equation}
\varphi(x,\tilde{x})=\sum_{n=-\infty}^{\infty}\sum_{\tilde{n}=-\infty}^{\infty}
\varphi_{n\tilde{n}}(x^i,\tilde{x}_i)
\, e^{i n x^1+ i
\tilde{n}\tilde{x}_1}\,,
\end{equation}
which satisfies $\varphi(x^1,\tilde{x}_1)=\varphi(x^1+2\pi,\tilde{x}_1+2\pi)$.
Thus, $n$ and $\tilde{n}$ are integers.
Then,
the action becomes $\int dx^id\tilde{x}^i {\cal L}_\flat$, where
\begin{eqnarray}
{\cal L}_\flat&\propto&\sqrt{\hat{g}}e^{-2\phi}\sum_{n,\tilde{n}}\Big[
-\frac{1}{2}\hat{g}^{ij}\{(\partial_i-i  n A_i-i \tilde{n}
B_i)\varphi_{n\tilde{n}}\}^*(\partial_j-i  n A_j-i \tilde{n}
B_j)\varphi_{n\tilde{n}}\nonumber\\ &
&\qquad-\frac{1}{2}\left[m^2+\left(\frac{1}{R}n+R\tilde{n}\right)^2\right]
|\varphi_{n\tilde{n}}|^2\Big]\,.
\end{eqnarray}

Now, we find the KK tower of the mass spectrum, which is very similar to the
left-mover mass operator in string theory with compactification \cite{Lust}. In
particular, there are infinite zero mode scalars in the case of
$m=0$ and $R=1$. Note that we did not assume the condition for massless field in
DFT, such as
$\tilde{\partial}^a=0$ or $\partial_a\tilde{\partial}^a=0$.

The ``level matching condition''  $\partial_a\tilde{\partial}^a=0$ was introduced into
DFT to have the symmetry of the massless background field described by a novel algebra
\cite{2,3,4,5,H,H2}.
The massive scalar field introduced here originates only from the symmetry of
the given fixed background fields such as in \cite{D2}. The validity of the
so-called level matching condition on the field theory of the scalar is not yet
clear.

If the condition $\partial_a\tilde{\partial}^a=0$ or equivalently
$\partial_A{\partial}^A=\eta^{AB}\partial_A\partial_B=0$ is imposed on the action of
the scalar field, the action can be expressed as
\begin{eqnarray}
S_\flat&=&\int dx d\tilde{x}\, e^{-2d}
\left(-\frac{\eta^{AB}+{\cal H}^{AB}}{2}\partial_A\varphi 
\partial_B\varphi-
\frac{1}{2}m^2\varphi^2\right)\nonumber \\\Rightarrow
S_\sharp&=&\int dx d\tilde{x}\, e^{-2d}
\left(-\frac{1}{2}{\cal H}^{AB}\partial_A\varphi 
\partial_B\varphi-
\frac{1}{2}m^2\varphi^2\right)
\,,
\end{eqnarray}
which is the natural form for the action. It should be noted that this is possibly not
the precise action, because the condition $\partial_a\tilde{\partial}^a=0$ should be
treated as a constraint on the field.

As for the KK mass spectrum discussed above, the condition becomes just
$n\tilde{n}=0$, and the mass spectrum can be derived from the action $S_\sharp$. Then,
the spectrum is
\begin{equation}
m^2+\frac{n^2}{R^2}\,,\quad m^2+R^2\tilde{n}^2\,,
\label{KKM}
\end{equation}
and there is a single state for the zero mode if $m=0$.

To briefly summarize the discussion thus far, the action $S_\flat$ leads to a rather
string-like spectrum for KK states but without level matching, whereas the action
$S_\sharp$ leads to the ``physical'' KK mass spectrum (with no stringy excitation
mode).

Now, we consider two-dimensional toroidal compactification and the KK mass spectrum.
The ansatz for the form of the scalar field is
\begin{equation}
\varphi(x,\tilde{x})=\sum_{n_1,n_2,\tilde{n}_1,\tilde{n}_2}
\varphi_{n_1n_2\tilde{n}_1\tilde{n}_2}(x^i,\tilde{x}_i)
\, e^{i n_1 x^1+i n_2 x^2+ i
\tilde{n}_1\tilde{x}_1+i \tilde{n}_2\tilde{x}_2}
\end{equation}
and for background geometry is
\begin{eqnarray}
& &g_{\mu\nu}=\left(
\begin{array}{ccc}
R^2 & R^2 \tau_1 & 0   \\
R^2 \tau_1 & R^2|\tau|^2 & 0   \\
0 & 0
& \hat{g}_{ij}
\end{array}
\right)\,,\\
& &g^{\mu\nu}=
{\footnotesize 
\left(
\begin{array}{ccc}
R^{-2}\tau_2^{-2}|\tau|^2& 
-R^{-2}\tau_2^{-2}\tau_1 & 0\\
-R^{-2}\tau_2^{-2}\tau_1 & R^{-2}\tau_2^{-2} &
0 \\ 0 & 0 &
\hat{g}^{ij} 
\end{array}
\right)
}
\,,
\end{eqnarray}
where we omit the vector degrees of freedom.
For the antisymmetric field, we assume that only the non-zero element is
$B\equiv b_{12}$.

We examine two cases. For the theory described by the action $S_\flat$, or without
the condition $\partial_A\partial^A=0$,
the mass spectrum is found to be
\begin{equation}
m^2+\left[\frac{1}{R\tau_2}\left\{n_2+B\tilde{n}_1-\tau_1
\left(n_1-B\tilde{n}_2\right)\right\}+R\tau_2\tilde{n}_2\right]^2
+\left[\frac{1}{R}(n_1-B\tilde{n}_2)+R(\tilde{n}_1+\tau_1\tilde{n}_2)\right]^2\,,
\end{equation}
as the sequence of mass squared.
On the other hand, for the theory described by the action $S_\sharp$, or with
the condition $\partial_A\partial^A=0$,
the mass spectrum is found to be
\begin{equation}
m^2+\frac{1}{R^2\tau_2^2}\left[\left\{n_2+B\tilde{n}_1-\tau_1
\left(n_1-B\tilde{n}_2\right)\right\}^2+\tau_2^2(n_1-B\tilde{n}_2)^2\right]
+R^2\left[(\tilde{n}_1+\tau_1\tilde{n}_2)^2+\tau_2^2\tilde{n}_2^2\right]\,,
\end{equation}
with $n_1\tilde{n}_1+n_2\tilde{n}_2=0$.
The spectrum includes the following cases:

For $\tilde{n}_1=\tilde{n}_2=0$,
\begin{equation}
m^2+\frac{1}{R^2\tau_2^2}\left[\left(n_2-\tau_1
n_1\right)^2+\tau_2^2n_1^2\right]
\,.
\end{equation}

For ${n}_1={n}_2=0$,
\begin{equation}
m^2+\left(\frac{B^2}{R^2\tau_2^2}
+R^2\right)\left[(\tilde{n}_1+\tau_1\tilde{n}_2)^2+\tau_2^2\tilde{n}_2^2\right]\,.
\end{equation}

For $\tilde{n}_1={n}_2=0$,
\begin{equation}
m^2+\frac{|\tau|^2}{R^2\tau_2^2}(n_1-B\tilde{n}_2)^2
+R^2|\tau|^2\tilde{n}_2^2\,.
\end{equation}

For ${n}_1=\tilde{n}_2=0$,
\begin{equation}
m^2+\frac{1}{R^2\tau_2^2}\left(n_2+B\tilde{n}_1\right)^2
+R^2\tilde{n}_1^2\,.
\end{equation}

In this section, we have examined the scalar field theory in the DFT background
fields. Because of the duality in the background fields,
the KK spectrum has a rich structure similar to that for the string theory with
toroidal compactification.

\section{a simply modified cosmological model
\label{ccos}}
Inspired by the success in the Hamiltonian analysis for the geodesic flow in the DFT
background, we apply a similar method to a modified model for
cosmology, which is related to the string cosmology or
pre-big bang models \cite{GV} and, thus, also related to DFT.

In DFT, the field is not doubled but the coordinate appears to be doubled.
The coordinate is of course not really doubled, since the level matching condition
and $O(D,D)$ rotation leads to $\tilde{\partial}^a=0$.

In the model here, we consider two metrics, $g$ and $\tilde{g}$.
Though our model describes a so-called bi-metric or bi-gravity theory, the degree of
freedom is to be mildly restricted. The interesting points from a physical perspective
are as follows. First, the model contains no inverse metric explicitly, which
generally leads to an infinite series of perturbations in the metric.
Second, we choose a model with no higher derivatives so as to abandon the perfect
elimination of extra degrees of freedom. Third, we expect that the extra degrees of
freedom in the metric can affect the resolution of cosmological problems.

For simplicity of the discussion, we consider $b_{\mu\nu}=0$.
The T-duality is expressed as a symmetry under $g_{\mu\nu}\rightarrow g^{\mu\nu}$.
We wish to consider a new metric $\tilde{g}$ and some moderate constraint to obtain
$\tilde{g} g=1$ approximately. At the same time, we consider that $\tilde{g} g= 1$
does not hold strictly at the very beginning of our universe.

If the metric is really doubled, 
\begin{equation}
P_{AB}\equiv\left(
\begin{array}{cc}
\tilde{g} & 1\\
1 & g
\end{array}\right)
\end{equation}
is not an exact projection matrix.
Instead of adopting the strong constraint $P^2=P$, we take 
\begin{equation}
P\dot{\cal H}P=0\,,
\end{equation}
as a constraint where
\begin{equation}
{\cal H}_{AB}\equiv\left(
\begin{array}{cc}
\tilde{g} & 0\\
0 & g
\end{array}\right)\,;
\end{equation}
and dot denotes the derivative with respect to the canonical time.%
\footnote{Please remember that we are inspired by the canonical formalism for
equations.}

Of course, $P^2=P$ implies $P\dot{\cal H}P=0$, but the inverse does not always hold.
Meanwhile, we expect the ``relaxation'' to $P^2=P$ to depend on the dynamics.

To avoid treating infinite degrees of freedom and complicated canonical decomposition
of the fields,   we consider cosmological settings in the present paper.
In other words, we consider a modification of the model shown in the last part of
Sec.~\ref{pca}. That is, we consider only the standard dynamics of finite degrees of
freedom.

The action we first consider is
\begin{equation}
S\rightarrow -\int dt\, e^{-\Phi}\left[\frac{1}{8}
\Tr(\dot{M}\eta\dot{M}\eta)+\dot{\Phi}^2+V\right]\,,
\end{equation}
where we add the constant potential $V$ to the Lagrangian.
We have an overall coefficient by the constant shift and redefinition of $\Phi$. Then
the action is rewritten as
\begin{equation}
S= -\frac{\lambda_s}{2}\int dt e^{-\Phi}\left[\frac{1}{8}
\Tr(\dot{M}\eta\dot{M}\eta)+\dot{\Phi}^2+V\right]\,,
\end{equation}
where $\lambda_s$ is the constant that represents the scale of string theory
\cite{GV}. Further, we use a new time parameter with
$dt=e^{-\Phi}d\tau$
\begin{equation}
S= -\frac{\lambda_s}{2}\int d\tau\left[\frac{1}{8}
\Tr(M'\eta M'\eta)+{\Phi'}^2+ e^{-2\Phi}V\right]\,,
\label{spbb}
\end{equation}
where the prime denotes differentiation with respect to $\tau$.

We now define the ``pseudo''-projection matrices
\begin{equation}
P=\frac{\eta+M}{2}\,,\qquad \bar{P}=\frac{\eta-M}{2}\,,
\end{equation}
and we wish to enforce
\begin{equation}
P{M}'P=\bar{P}{M}'\bar{P}=0\,,
\end{equation}
using some constraints.
Please note that thus far the ``problem'' is very similar to the case of the
particle motion in the DFT background.

Now, the Lagrangian $L_\Lambda$ with the constraint term is
\begin{equation}
L_\Lambda=\frac{\lambda_s}{2}\left[-\frac{1}{8}
{M}'^{AB}{M}'_{AB}+\bar{\Lambda}_{AB}\bar{P}^{AC}{M}'_{CD}\bar{P}^{DB}
+\Lambda_{AB}{P}^{AC}{M}'_{CD}{P}^{DB}-{\Phi}'^2-e^{-2\Phi}V\right]\,,
\end{equation}
where $A, B$ range over $1,\ldots , D$.
It must be noted that the constraint terms are invariant with respect to the choice of
time in the action.

The conjugate momentum for $M$ is derived as
\begin{equation}
\Pi^{AB}=\frac{\partial L_\Lambda}{\partial
M'_{AB}}=\frac{\lambda_s}{2}\left[-\frac{1}{4}
{M}'^{AB}+\bar{P}^{AC}\bar{\Lambda}_{CD}\bar{P}^{DB}
+{P}^{AC}\Lambda_{CD}{P}^{DB}\right]\,,
\end{equation}
whereas the momentum for $\Phi$ is 
\begin{equation}
\Pi_{\Phi}=\frac{\partial L_\Lambda}{\partial {\Phi'}}=-\lambda_s {\Phi'}\,.
\end{equation}
Therefore, the Hamiltonian of the system becomes
\begin{eqnarray}
H_\Lambda&=&\Pi^{AB}{M}'_{AB}+\Pi_{\Phi}{\Phi}'-L_\Lambda\nonumber \\
&=&-\frac{4}{\lambda_s}
\left[\Pi^{AB}-\frac{\lambda_s}{2}\left(\bar{P}^{AC}\bar{\Lambda}_{CD}\bar{P}^{DB}
+{P}^{AC}\Lambda_{CD}{P}^{DB}\right)\right]^2-\frac{1}{2\lambda_s}\Pi_{\Phi}^2
+\frac{\lambda_s}{2}e^{-2\Phi}V\,,
\end{eqnarray}
where $({\cal M}^{AB})^2$ means ${\cal M}^{AB}{\cal M}_{AB}$.

Finding an exact solution for the Lagrange multiplier is difficult or needs
more recursive constraints. We consider simplification by using the assumed relation,
$P^2\simeq P$ and $\bar{P}^2\simeq\bar{P}$.
In this section, the symbol $\simeq$ is used to indicate this assumed approximation
adopted by us.

Then, we can express the following relations
\begin{equation}
\frac{\partial H_\Lambda}{\partial\bar{\Lambda}_{AB}}\simeq 2\bar{P}^{AC}
\left(\Pi_{CD}-\frac{\lambda_s}{2}\,\bar{\Lambda}_{CD}\right)\bar{P}^{DB}= 0\,,
\end{equation}
and
\begin{equation}
\frac{\partial H_\Lambda}{\partial\Lambda_{AB}}\simeq 2P^{AC}
\left(\Pi_{CD}-\frac{\lambda_s}{2}\,\Lambda_{CD}\right)P^{DB}= 0\,.
\end{equation}
The solutions of these equations are
\begin{equation}
\frac{\lambda_s}{2}\,\bar{\Lambda}_{AB}\simeq\Pi_{AB}+P_{AC}L_1^{CD}\bar{P}_{DB}
+\bar{P}_{AC}L_2^{CD}{P}_{DB}+P_{AC}L_3^{CD}P_{DB}\,,
\end{equation}
and
\begin{equation}
\frac{\lambda_s}{2}\,\Lambda_{AB}\simeq\Pi_{AB}+P_{AC}\bar{L}_1^{CD}\bar{P}_{DB}+
\bar{P}_{AC}\bar{L}_2^{CD}{P}_{DB}+\bar{P}_{AC}\bar{L}_3^{CD}\bar{P}_{DB}\,,
\end{equation}
where $L_i$ and $\bar{L}_i$ $(i=1,2,3)$ are arbitrary tensors.

Substituting these into the original Hamiltonian $H_\Lambda$, finally, we obtain the
Hamiltonian in which the multipliers are eliminated as
\begin{eqnarray}
H_\Lambda&\simeq&-\frac{4}{\lambda_s}
\left(\Pi^{AB}-\bar{P}^{AC}\Pi_{CD}\bar{P}^{DB}
-{P}^{AC}\Pi_{CD}{P}^{DB}\right)^2-\frac{1}{2\lambda_s}\Pi_{\Phi}^2
+\frac{\lambda_s}{2}e^{-2\Phi}V\nonumber
\\ &\simeq&-\frac{8}{\lambda_s}
\Pi^{AB}{P}_{BC}\Pi^{CD}\bar{P}_{DA}-\frac{1}{2\lambda_s}\Pi_{\Phi}^2
+\frac{\lambda_s}{2}e^{-2\Phi}V\,.
\end{eqnarray}

We consider this as a new Hamiltonian
\begin{equation}
H_*\equiv -\frac{8}{\lambda_s}
\Pi^{AB}{P}_{BC}\Pi^{CD}\bar{P}_{DA}-\frac{1}{2\lambda_s}\Pi_{\Phi}^2
+\frac{\lambda_s}{2}e^{-2\Phi}V\,.
\end{equation}
The Hamilton equation thus becomes
\begin{equation}
{M}'_{AB}=\frac{\partial H_*}{\partial \Pi^{AB}}=
-\frac{8}{\lambda_s}
\left(\bar{P}_{AC}\Pi^{CD}{P}_{DB}
+{P}_{AC}\Pi^{CD}\bar{P}_{DB}\right)\,.
\end{equation}
Therefore,
$P{M'}P\simeq 0$ and $\bar{P}{M'}\bar{P}\simeq 0$ are verified.

The Wheeler-DeWitt equation can be obtained by the replacement
\begin{equation}
\Pi^{AB}\rightarrow\hat{\Pi}^{AB}=-i\,\frac{\delta}{\delta M_{AB}}\,,\quad
\Pi_{\Phi}\rightarrow\hat{\Pi}_{\Phi}=-i\,\frac{\delta}{\delta \Phi}\,,
\end{equation}
in the Hamiltonian and introducing the wave function $\Psi$ satisfying
\begin{equation}
\hat{H}_*\,\Psi(M, \Phi)=0\,,
\end{equation}
modulo ordering of operators.

\section{``minisuperspace'' version of the modified model}
\label{ms}
In this section, we examine the previous procedure of modification in
the minisuperspace model.
We suppose
\begin{equation}
M_{AB}=\left(
\begin{array}{cc}
\tilde{A}(\tau)\delta^{ab} & 0\\
0 & A(\tau)\delta_{\mu\nu}
\end{array}\right)\,.
\end{equation}
Since the kinetic part of $M$ of the Lagrangian is
\begin{equation}
L_0=-\frac{\lambda_s}{16}
{M}'^{AB}{M}'_{AB}=-\frac{\lambda_s D}{8}A'\tilde{A}'\,,
\end{equation}
the conjugate momenta of $A$ and $\tilde{A}$ are
\begin{equation}
\pi\equiv\frac{\partial L_0}{\partial A'}=-\frac{\lambda_s D}{8}\tilde{A}'\,,\quad
\tilde{\pi}\equiv\frac{\partial L_0}{\partial \tilde{A}'}=-\frac{\lambda_s
D}{8}{A}'\,.
\end{equation}
Thus the naive kinetic part of the Hamiltonian becomes
\begin{equation}
H_0=-\frac{8}{\lambda_s D}\pi\tilde{\pi}=-\frac{4}{\lambda_s D}
(\pi\tilde{\pi}+\tilde{\pi}\pi)\,.
\end{equation}

To derive the modified model, it is useful to rewrite the above Hamiltonian as
\begin{equation}
H_0=-\frac{4}{\lambda_s D}\,\Tr\left[
\left(\begin{array}{cc}0&1\\1&0\end{array}\right)
\left(\begin{array}{cc}\tilde{\pi}&0\\0&\pi\end{array}\right)
\left(\begin{array}{cc}0&1\\1&0\end{array}\right)
\left(\begin{array}{cc}\tilde{\pi}&0\\0&\pi\end{array}\right)
\right]\,.
\end{equation}
In this case, the modification explained in the previous section becomes
\begin{eqnarray}
H_{*0}&=&-\frac{2}{\lambda_s D}\,\Tr\left[
\left(\begin{array}{cc}-\tilde{A}&1\\1&-A\end{array}\right)
\left(\begin{array}{cc}\tilde{\pi}&0\\0&\pi\end{array}\right)
\left(\begin{array}{cc}\tilde{A}&1\\1&A\end{array}\right)
\left(\begin{array}{cc}\tilde{\pi}&0\\0&\pi\end{array}\right)
\right]\nonumber \\
&=&-\frac{2}{\lambda_s D}\,(\pi\tilde{\pi}+\tilde{\pi}\pi-A\pi
A\pi-\tilde{A}\tilde{\pi}\tilde{A}\tilde{\pi})\,.
\end{eqnarray}
The Hamiltonian for the minisuperspace version of our modified model is
\begin{equation}
H_*=-\frac{2}{\lambda_s D}\,(\pi\tilde{\pi}+\tilde{\pi}\pi-A\pi
A\pi-\tilde{A}\tilde{\pi}\tilde{A}\tilde{\pi})-\frac{1}{2\lambda_s}\Pi_{\Phi}^2
+\frac{\lambda_s}{2}e^{-2\Phi}V\,.
\label{mssH}
\end{equation}

The Hamilton equations in the case with $V\equiv constant$ are found to be
\begin{equation}
A'=\frac{\partial H_*}{\partial \pi}=-\frac{4}{\lambda_s D}(\tilde{\pi}-A\pi A)\,,
\label{eq1}
\end{equation}
\begin{equation}
\tilde{A}'=\frac{\partial H_*}{\partial \tilde{\pi}}=-\frac{4}{\lambda_s
D}({\pi}-\tilde{A}\tilde{\pi}\tilde{A})\,,
\label{eq2}
\end{equation}
\begin{equation}
\pi'=-\frac{\partial H_*}{\partial {A}}=-\frac{4}{\lambda_s D}\pi A \pi\,,
\label{eq3}
\end{equation}
\begin{equation}
\tilde{\pi}'=-\frac{\partial H_*}{\partial \tilde{A}}=-\frac{4}{\lambda_s
D}\tilde{\pi}\tilde{A}\tilde{\pi}\,,
\label{eq4}
\end{equation}
and
\begin{equation}
{\Phi}'=\frac{\partial H_*}{\partial \Pi_\Phi}=-\frac{\Pi_\Phi}{\lambda_s}\,,\qquad
{\Pi'_\Phi}=-\frac{\partial H_*}{\partial \Phi}=\lambda_s V e^{-2\Phi}\,.
\label{eq5}
\end{equation}

A special solution can be found for these equations. 
First, $\pi\equiv 0$ is found to be a solution for (\ref{eq3}).
Then, for (\ref{eq2}) and (\ref{eq4}), we find that
$\tilde{A}\tilde{\pi}=-\frac{\lambda_s \sqrt{D}}{2}C= constant$ is a solution.
Now, the solution is
\begin{equation}
\tilde{A}(\tau)=\frac{1}{A_0}\exp\left[-\frac{2}{\sqrt{D}}C(\tau-\tau_0)\right]\,,\quad
{A}(\tau)={A_0}\exp\left[\frac{2}{\sqrt{D}}C(\tau-\tau_0)\right]+\delta\,,
\end{equation}
where $A_0$, $\tau_0$, and $\delta$ are constants.
Then, the solution for $\Phi$ is given by
\begin{equation}
\Phi(\tau)=C(\tau-\tau_0)\qquad {\rm for}\quad V=0\,,
\end{equation}
\begin{equation}
\Phi(\tau)=\ln \left[\frac{\sqrt{V}}{C}\sinh C(\tau-\tau_0)\right]\qquad {\rm
for}\quad V>0\,,
\end{equation}
when the Hamiltonian constraint $H_*\approx 0$ is taken into consideration.
The similarity to the known string cosmological solution \cite{GV} is obvious, up to
the possible constant deviation $\delta$ in $A$.
Note that here we use the ``time'' $\tau$, and the original temporal parameter $t$ is
given by
$t=\int e^{-\Phi}d\tau$. For readers' convenience, we review the usual
string-cosmological solution as a function of $\tau$ in Appendix \ref{appA}.

For the solution, we find that $A\tilde{A}\rightarrow 1$ when $\tau\rightarrow
+\infty$. Of course, the solution exists for the case of time reversal as well.
In order to determine the condition at which the deviation from
$A\tilde{A}=1$ is significant, some initial conditions must be considered. 
In the next
section, we consider quantum cosmology of our model in the minisuperspace.

\section{Quantum Cosmology
\label{QC}}
Quantum cosmological treatment of the string cosmology has been widely studied
\cite{qsc}. In our model, we can obtain the minisuperspace Wheeler-DeWitt
equation by replacing
$\pi\rightarrow -i\frac{\partial}{\partial A}$, $\tilde{\pi}\rightarrow
-i\frac{\partial}{\partial \tilde{A}}$, and $\Pi_\Phi\rightarrow
-i\frac{\partial}{\partial
\Phi}$ in (\ref{mssH}).
Thus, we obtain
\begin{equation}
\left[\frac{2}{\lambda_s
D}\,\left(2\frac{\partial}{\partial A}\frac{\partial}{\partial
\tilde{A}}-A\frac{\partial}{\partial A}
A\frac{\partial}{\partial
A}-\tilde{A}\frac{\partial}{\partial
\tilde{A}}\tilde{A}\frac{\partial}{\partial
\tilde{A}}\right)+\frac{1}{2\lambda_s}
\frac{\partial^2}{\partial\Phi^2}
+\frac{\lambda_s}{2}e^{-2\Phi}V\right]\Psi=0\,,
\end{equation}
where $\Psi$ is the wave function of the universe.%
\footnote{In this toy model, we can easily neglect the operator ordering.}

Using the following variables, we can mostly simplify the above equation.
\begin{equation}
x=\frac{\sqrt{D}}{4}\ln A/\tilde{A}\,,\qquad
y=\frac{\sqrt{D}}{4}\ln A\tilde{A}\,.
\end{equation}
Up to the ordering, we have 
\begin{equation}
\left[-\frac{1+e^{-({4}/{\sqrt{D}})y}}{2}\frac{\partial^2}{\partial
x^2}-\frac{\partial}{\partial
y}\frac{1-e^{-({4}/{\sqrt{D}})y}}{2}\frac{\partial}{\partial y}+
\frac{\partial^2}{\partial\Phi^2}
+{\lambda_s^2}e^{-2\Phi}V\right]\Psi=0\,.
\end{equation}
Now, we consider the case with $V=constant$.

If we assume a solution with multiplicative form,
\begin{equation}
\psi(x,y,\Phi)=X(x)Y(y)Z(\Phi)\,,
\end{equation}
then the differential equation becomes
\begin{eqnarray}
-\frac{1+e^{-({4}/{\sqrt{D}})y}}{2}\frac{1}{X(x)}\frac{\partial^2
X(x)}{\partial x^2}&-&\frac{1}{Y(y)}\frac{\partial}{\partial
y}\frac{1-e^{-({4}/{\sqrt{D}})y}}{2}\frac{\partial Y(y)}{\partial y}\nonumber\\
&&\qquad\qquad+
\frac{1}{Z(\Phi)}\frac{\partial^2 Z(\Phi)}{\partial\Phi^2}
+{\lambda_s^2}e^{-2\Phi}V=0
\end{eqnarray}
and can be separated as
\begin{equation}
-\frac{\partial^2
X_k(x)}{\partial x^2}=k^2X_k(x)\,,\quad -\frac{\partial^2 Z_K(\Phi)}{\partial\Phi^2}
-{\lambda_s^2}e^{-2\Phi}VZ_K(\Phi)-K^2Z_K(\Phi)=0\,,
\label{se}
\end{equation}
and
\begin{equation}
-\frac{\partial}{\partial
y}\frac{1-e^{-({4}/{\sqrt{D}})y}}{2}\frac{\partial Y_{kK}(y)}{\partial
y}+\left(-K^2+
\frac{1+e^{-({4}/{\sqrt{D}})y}}{2}k^2\right)Y_{kK}(y)=0\,.
\label{yyy}
\end{equation}

The non singular real solution of (\ref{yyy}) at $y=0$ can be found to be
{\footnotesize
\begin{equation}
Y_{kK}(y)=e^{-\sqrt{k^2-2K^2}y}F\left(
{
\frac{1+\sqrt{1-k^2}+\sqrt{k^2-2K^2}}{2},
\frac{1-\sqrt{1-k^2}+\sqrt{k^2-2K^2}}{2},1;1-e^{-2y}
}
\right)\,,
\label{ohh}
\end{equation}
}
where $F(\alpha,\beta,\gamma; z)$ is the Gauss' hypergeometric function.%
\footnote{Equivalent expressions for the solution are exhibited in Appendix \ref{bb}.}
The solutions for (\ref{se}) can be obtained from the standard quantum cosmological
model reviewed in Appendix \ref{appA}.

If $K=\pm k$, $Y_{k\,\pm k}(y)$ has a maximum at $y=0$. When we construct a
wave packet for the cosmological wave function \cite{qsc}, the peak of this wave
packet in terms of parameter $y$ is naturally located at $y=0$. This is because there
is no other fixed peak apart from $y=0$ in the constituent waves and waves
destructively interfere with each other at
$y\neq 0$. Thus, the approximate scale factor duality $A\tilde{A}\simeq 1$
is expected even at the ``beginning'' of the quantum universe.
The detailed investigation on the behavior of the universe is left for future
research.

\section{Summary and outlook
\label{summary}}
The present paper consists of two parts. In the first part (Sec.
\ref{ems}-\ref{part}), the motion of the particle in the background field in DFT has
been investigated. 
We have explicitly shown that the geodesic in the $2D$-dimensional doubled-spacetime
cannot  be the geodesic in the $D$-dimensional spacetime. The
geodesic equation in the $D$-dimensional spacetime is derived by the
Hamiltonian formalism of the constraint system and is found to be the geodesic
flow equation. 
The KK excited states have been studied and are found to be the
same spectra in string theory but without stringy excitations.

In the second part (Sec.
\ref{ccos}-\ref{QC}), we have considered the string cosmology with a bimetric model 
inspired by the constraint method discussed in the first part.
Our method for the restriction on the metrics functions well, at least in the present
reduced model for cosmology.

The canonical formalism for the model with two metrics should be further studied
as the field theory of gravity. Of course, the important consideration is the
deviation from the precise duality in the model. It will be interesting to find
whether the extra degrees of freedom affects the cosmology and gravitational
theory. We wish to apply our approach to other systems with some duality or
other symmetries.

\acknowledgments
The authors thank the Yukawa Institute for Theoretical Physics at Kyoto University.
Discussions during the YITP workshop YITP-W-11-05 on ``Field Theory and String
Theory'' were useful to complete this work.

We thank David Berman for providing information on his seminal work on the beta
function in the doubled formalism \cite{berman1} and papers on generalizing the
doubled formalism to M-theory \cite{berman2}, which are related to the more recent
development around DFT.


\begin{appendix}

\section{Classical and quantum equations in the standard string
cosmology (pre-big bang) model
\label{appA}}
In this Appendix, we review the the standard string-cosmological solution
\cite{GV,qsc} using our notation for convenience.
First, we define
\begin{equation}
M_{AB}=\left(
\begin{array}{cc}
{A^{-1}}(\tau)\delta^{ab} & 0\\
0 & A(\tau)\delta_{\mu\nu}
\end{array}\right)\,,
\end{equation}
where $A$ is the square of the scale factor.
Then, the action (\ref{spbb}) is expressed as
\begin{equation}
S= -\frac{\lambda_s}{2}\int d\tau\left[-\frac{D}{4}
\left(\frac{A'}{A}\right)^2+{\Phi'}^2+ e^{-2\Phi}V\right]\equiv\int L d\tau\,.
\end{equation}
The conjugate momentum for $A$ and $\Phi$ is given by
\begin{equation}
\Pi\equiv\frac{\partial L}{\partial A'}=\frac{\lambda_s D}{4}\frac{A'}{A^2}\quad
{\rm and}\quad
\Pi_\Phi\equiv-\lambda_s\Phi'\,,
\end{equation}
respectively.
Now, we obtain the Hamiltonian as
\begin{equation}
H\equiv\Pi A'+\Pi_\Phi\Phi'-L=\frac{2}{\lambda_s
D}A\Pi A\Pi-\frac{1}{2\lambda_s}\Pi_\Phi^2+\frac{\lambda_s}{2}
e^{-2\Phi}V\,.
\end{equation}
For the case with $V=constant$, the Hamilton equations are
\begin{equation}
A'=\frac{\partial H}{\partial \Pi}=\frac{4}{\lambda_s D}A\Pi A\,,
\qquad
\Pi'=-\frac{\partial H}{\partial A}=-\frac{4}{\lambda_s D}\Pi A \Pi\,,
\label{AAA}
\end{equation}
\begin{equation}
\Phi'=\frac{\partial H}{\partial \Pi_{\Phi}}=-\frac{\Pi_{\Phi}}{\lambda_s}\,,\qquad
\Pi_{\Phi}'=-\frac{\partial H}{\partial \Phi}=\lambda_s V e^{-2\Phi}\,,
\end{equation}
From (\ref{AAA}), we have
\begin{equation}
A\Pi=\frac{\lambda_s \sqrt{D}}{2}C\,,
\end{equation}
and we find the solution
\begin{equation}
A(\tau)=A_0\exp \left[\frac{2}{\sqrt{D}}C(\tau-\tau_0)\right]\,,
\end{equation}
where $C$, $A_0$, and $\tau_0$ are integration constants.
Then, the solution for $\Phi$ is given by
\begin{equation}
\Phi(\tau)=C(\tau-\tau_0)\qquad {\rm for}\quad V=0\,,
\end{equation}
\begin{equation}
\Phi(\tau)=\ln \left[\frac{\sqrt{V}}{C}\sinh C(\tau-\tau_0)\right]\qquad {\rm
for}\quad V>0\,.
\end{equation}

The Wheeler-DeWitt equation for the standard quantum string cosmology is \cite{qsc}
\begin{equation}
\left[-\frac{2}{\lambda_s
D}A\frac{\partial}{\partial
A}
A\frac{\partial}{\partial
A}+\frac{1}{2\lambda_s}\frac{\partial^2}{\partial\Phi^2}+\frac{\lambda_s}{2}
e^{-2\Phi}V\right]\Psi=0\,.
\end{equation}
By using the variable $x=\frac{\sqrt{D}}{2}\ln A$, we rewrite the equation as
\begin{equation}
\left[-\frac{\partial^2}{\partial
x^2}+\frac{\partial^2}{\partial\Phi^2}+{\lambda_s^2}
e^{-2\Phi}V\right]\Psi(x,\Phi)=0\,.
\end{equation}
The general separable solution for a constant $V$ is found to be
\begin{equation}
\left(\frac{\lambda_s\sqrt{V}}{2}\right)^{\pm ik}\Gamma(1\mp ik)\,
J_{\mp
ik}(\lambda_s\sqrt{V}e^{-\Phi})\stackrel{\Phi\rightarrow+\infty}{\longrightarrow}
e^{\pm ik\Phi-ikx}\,.
\end{equation}

\section{Other expressions for the solution (\ref{ohh})
\label{bb}}

The solution (\ref{ohh}) can be expressed in the following forms:
{\footnotesize
\begin{eqnarray}
Y_{kK}(y)&=&e^{-\sqrt{k^2-2K^2}y}\,F\left(
{
\frac{1+\sqrt{1-k^2}+\sqrt{k^2-2K^2}}{2},
\frac{1-\sqrt{1-k^2}+\sqrt{k^2-2K^2}}{2},1;1-e^{-2y}
}
\right)\nonumber \\
&=&e^{+\sqrt{k^2-2K^2}y}\,F\left(
{
\frac{1+\sqrt{1-k^2}-\sqrt{k^2-2K^2}}{2},
\frac{1-\sqrt{1-k^2}-\sqrt{k^2-2K^2}}{2},1;1-e^{-2y}
}
\right)\nonumber \\&=&e^{(1-\sqrt{1-k^2})y}\,F\left(
{
\frac{1-\sqrt{1-k^2}-\sqrt{k^2-2K^2}}{2},
\frac{1-\sqrt{1-k^2}+\sqrt{k^2-2K^2}}{2},1;1-e^{2y}
}
\right)\nonumber \\&=&e^{(1+\sqrt{1-k^2})y}\,F\left(
{
\frac{1+\sqrt{1-k^2}-\sqrt{k^2-2K^2}}{2},
\frac{1+\sqrt{1-k^2}+\sqrt{k^2-2K^2}}{2},1;1-e^{2y}
}
\right)\nonumber \\&=&e^{(1-\sqrt{1-k^2})y}\,F\left(
{
\frac{1-\sqrt{1-k^2}-\sqrt{k^2-2K^2}}{2},
\frac{1-\sqrt{1-k^2}+\sqrt{k^2-2K^2}}{2},1-\sqrt{1-k^2};e^{2y}
}
\right)\nonumber \\&=&e^{(1+\sqrt{1-k^2})y}\,F\left(
{
\frac{1+\sqrt{1-k^2}-\sqrt{k^2-2K^2}}{2},
\frac{1+\sqrt{1-k^2}+\sqrt{k^2-2K^2}}{2},1+\sqrt{1-k^2};e^{2y}
}
\right)\,.\nonumber \\
& &
\end{eqnarray}
}

\end{appendix}


\bibliographystyle{apsrev4-1}

\end{document}